\documentclass[12pt]{article}
\usepackage[utf8]{inputenc}
\usepackage{graphicx}
\usepackage{multicol}
\usepackage{placeins}
\usepackage{flafter}
\usepackage{caption}
\usepackage{subcaption}
\usepackage{amsmath, amsthm}
\usepackage{amsfonts}
\usepackage{amssymb}
\usepackage{url}
\usepackage{color}
\usepackage{pdfpages}
\usepackage{authblk}
\usepackage[dvipsnames]{xcolor}
\usepackage[mathlines]{lineno}
\usepackage{booktabs} %for making tables
\usepackage{titlesec} % to make subsubsubsections as paragraphs
\usepackage[toc,page]{appendix}
\usepackage{hyperref}
\newif\ifshowcomments
\showcommentsfalse
\newcommand{\tsup}{\operatorname{tsup}}
\newcommand{\rel}{\operatorname{rel}}
\newtheorem{example}{Example}

\begin{document}

\title{
A novel association and ranking approach identifies factors affecting educational outcomes of STEM majors}

\author{
Kira Adaricheva\textsuperscript{\rm 1},
Jonathan T. Brockman\textsuperscript{\rm 2},
Gillian Z.  Elston\textsuperscript{\rm 1},
Lawrence Hobbie\textsuperscript{\rm 3},
%Nathalia Holtzman\textsuperscript{\rm 4},
Skylar Homan\textsuperscript{\rm 4},
Mohamad Khalefa\textsuperscript{\rm 5},
Jiyun V. Kim\textsuperscript{\rm 6},
Rochelle K. Nelson\textsuperscript{\rm 7},
Sarah Samad\textsuperscript{\rm 4},
Oren Segal\textsuperscript{\rm 8}
}
\affil{
\textsuperscript{\rm 1}Departments of Mathematics, Hofstra University, Hempstead, NY 11549, USA.\\
\textsuperscript{\rm 2} 
Department of Chemistry, Suffolk County Community College, Selden, NY 11784, USA.\\
\textsuperscript{\rm 3} 
Department of Biology, Adelphi University, Garden City, NY 11530 USA.\\
\textsuperscript{\rm 4} 
Hofstra University, Hempstead, NY 11549, USA.\\
\textsuperscript{\rm 5} 
Mathematics, Computer \& Information Science, SUNY Old Westbury, NY 11568, USA.\\ 

\textsuperscript{\rm 6} 
Department of Biology, Hofstra University, Hempstead, NY 11549, USA.\\
\textsuperscript{\rm 7} 
Department of Biological Sciences and Geology, Queensborough Community College, Queens, 11364 NY, USA.\\
\textsuperscript{\rm 8}  
Department of Computer Science, Hofstra University, Hempstead, NY 11549, USA.\\
kira.adaricheva@hofstra.edu, brockmj@sunysuffolk.edu, 
gillian.elston@hofstra.edu, hobbie@adelphi.edu,
%nholtzman@qc.cuny.edu,
khalefam@oldwestbury.edu,
jiyun.kim@hofstra.edu,
rnelson@qcc.cuny.edu,
oren.segal@hofstra.edu

}

\date{
%\today
}

\maketitle

\abstract{
Background: Improving undergraduate success in STEM requires identifying actionable factors that impact student outcomes, allowing institutions to prioritize key leverage points for change. We examined academic, demographic, and institutional factors that might be associated with graduation rates at two four-year colleges in the northeastern United States using a novel association algorithm called $D$-basis to rank attributes associated with graduation. Importantly, the data analyzed included tracking data from the National Student Clearinghouse on students who left their original institutions to determine outcomes following transfer.\\ 

Results: Key predictors of successful graduation include performance in introductory STEM courses, the choice of first mathematics class, and flexibility in major selection. High grades in introductory biology, general chemistry, and mathematics courses were strongly correlated with graduation. At the same time, students who switched majors – especially from STEM to non-STEM – had higher overall graduation rates. Additionally, Pell eligibility and demographic factors, though less predictive overall, revealed disparities in time to graduation and retention rates.\\

Conclusions: The findings highlight the importance of early academic support in STEM gateway courses and the implementation of institutional policies that provide flexibility in major selection. Enhancing student success in introductory mathematics, biology, and chemistry courses could greatly influence graduation rates. Furthermore, customized mathematics pathways and focused support for STEM courses may assist institutions in optimizing student outcomes. This study offers data-driven insights to guide strategies to increase STEM degree completion.}
  

\tableofcontents
\section{Introduction}
Higher education has long been viewed as a valuable path to an economically secure life in the United States \cite{taylor2011college}. 
Long-term economic trends and widely publicized research on the value of a college education have increased the desire among parents for their children to earn college degrees, preferably in fields seen as leading most directly to desirable jobs, such as those in science, technology, engineering, and math (STEM) \cite{vuolo2016value}.
Indeed, the national percentage of earned STEM bachelor’s degrees has steadily been increasing, reaching one-third of all degrees awarded in 2018 \cite{fry2021stem}.

Higher education in STEM has multiple benefits for individuals and society. 
Individuals with STEM degrees often command higher starting salaries and improved job security. 
Over time, these STEM degree holders have greater access to promotion opportunities and earn substantially more than those with non-STEM degrees. 
On a societal level, STEM professionals are essential for continuous innovation and technological progress.
Employment growth in STEM fields also leads, through the economic multiplier effect, to job growth and economic activity in related sectors such as goods and services \cite{jobs2020stem}.
These considerations suggest that an effective and inclusive STEM education system could contribute to individual and social thriving.

\subsection{STEM education trends and disparities}
 
An earlier metaphor of STEM education as a ``pipeline" (or even a ``leaky pipeline") has been replaced by that of a braided river or multiple interconnected pathways \cite{Batchelor2021},
reflecting the observation that there are multiple entry and exit points.
For example, of those who entered four-year colleges nationwide in 2004 as STEM majors, only 24.7\% graduated with STEM degrees within 6 years and another 22.6\% completed a non-STEM degree, whereas 52\% did not graduate \cite{eagan2014examining}. 
Of the students in this cohort who graduated with a STEM degree, 17.7\% were not STEM majors originally.
Another study found that within three years of their initial enrollment, 35\% of all students who initially declared a STEM major changed their area of study compared to 29\% of those who declared non-STEM majors \cite{leu2017beginning}.

Latinx and Black students enroll in and complete STEM degree programs at a lower rate than their peers despite initial high interest in STEM degrees \cite{chen2013stem, asai2020race, riegle2019does,riegle2010who}. 
For example, in 2018 only 12\% of Latinx and 7\% of Black students earned a STEM bachelor’s degree.
The result is a STEM workforce that included only 8\% Latinx and 9\% Black workers, resulting in an overrepresentation of Asian and White students and workers~\cite{fry2021stem}. 
Latinx and Black students are thus starkly underrepresented in STEM degrees and jobs.

Under-representation of women also remains a challenge in some areas of STEM such as engineering, mathematics, and computer science, whereas in other areas such as biology the representation of women is close to equal at least in proportion of undergraduate majors (although not in the faculty) \cite{NCSES2023}. 
These racial, ethnic, and gender disparities reflect and help to perpetuate broader socioeconomic inequities in the United States. 
Improving STEM degree completion rates would enable thousands of students from disadvantaged groups to earn a STEM degree, enhancing their life prospects \cite{morrison2019college} and decreasing race/ethnic disparities in STEM education \cite{bottia2021factors}.

\vspace{0.3cm} 

\subsection{Research context}
 
Improving the success of undergraduates in STEM requires identifying actionable factors that impact student outcomes, allowing institutions to prioritize key leverage points for change. Many studies have examined factors related to the individual student (demography, high school preparation, and college experience), the institution, and students' perceptions of their educational environment \cite{campbell2022how}, \cite{hansen2023importance}, \cite{hatfield2022do}, \cite{meaders2020undergraduate}. These investigations have employed a variety of methodologies, ranging from qualitative interviews and surveys to quantitative statistical analyses. Some studies focus on small sample sizes at single institutions, while others use large national datasets to uncover trends. These studies highlight the many challenges students face. 

Considering these challenges, this study addresses the following research question: What academic, demographic, and institutional factors are most strongly associated with graduation rates at two four-year colleges in the northeastern United States? 
By identifying these factors, the current research aims to provide actionable insights that institutions can utilize to improve student outcomes, particularly within the STEM disciplines. 
Institutions can focus their resources on implementing effective strategies to support students, particularly those from historically underrepresented or economically disadvantaged backgrounds.

 \vspace{0.3cm}
 \subsection{Theoretical frameworks}
 
Astin’s (1993) Input-Environment-Output (I-E-O) model provides a valuable framework for examining factors that shape student outcomes in higher education \cite{Feldman94}, \cite{JohnGuer2014}. 
The model considers three key components: inputs, environment, and outputs. 
Inputs refer to the qualities that students bring with them when they enter an institution, such as their background, academic achievements, and personal characteristics. 
These inputs serve as starting points, influencing how students engage with the educational process.

In Astin’s model, the environment encompasses everything the student encounters within the institution that affects their educational experience \cite{Feldman94}, \cite{JohnGuer2014}. 
This includes formal academic experiences, such as coursework and faculty interactions, and the informal aspects of college life, such as peer relationships and extracurricular activities. 
Institutions play a central role in shaping the environment and determining the types of support systems and learning opportunities available to students, which can significantly influence their development and success.

Outputs, the final element in the model, represent the educational process's outcomes, which can include measures such as graduation rates, academic performance, and career readiness \cite{Feldman94}, \cite{JohnGuer2014}. The I-E-O model emphasizes the interaction between students' inputs, the environments they navigate, and the outputs that emerge. By examining these connections, educators and researchers can better understand how to support students and improve institutional effectiveness.

This framework considers the complex factors influencing student outcomes, particularly graduation rates. In the current study, the ‘inputs’ include student demographic characteristics (e.g., gender, race, and ethnicity), overall academic performance, and socioeconomic factors such as Pell eligibility. 
These inputs set the foundation for understanding the baseline factors that students bring into the higher education environment, which strongly correlate with their academic trajectories and graduation likelihood. 
By incorporating these variables, the I-E-O model allows for a nuanced analysis of how individual backgrounds affect student success, particularly in the diverse populations of the two four-year northeastern colleges examined in this study.

The “environment” in this manuscript focuses on the academic pathways that interact with the students’ inputs to influence their progression toward graduation. 
For example, the choice of initial STEM courses and the timing of the switch of majors are part of the “environment” that either facilitate or hinder student success. 
Astin's model emphasizes that student success is not solely based on their inputs but is also significantly shaped by the institutional environment, particularly the courses and academic support structures students engage with during their college experience.

Finally, the “outputs” analyzed in this study include whether a student graduates with a bachelors degree, whether they complete a degree in STEM or non-STEM fields, whether they graduate from their original institution or after transferring, and whether they graduate within the typical four-year timeframe. 
Using the I-E-O model, this study connects the academic, demographic, and institutional factors to student outcomes, identifying barriers and proposing targeted interventions. 
This holistic approach is essential for understanding how varying inputs and institutional environments affect graduation rates, offering actionable insights for improving student success at the two colleges under study.

 \vspace{0.3cm}

\subsection{Study overview}
 
The current study utilized a fine-grained analysis of individual student records from multiple years at two private four-year institutions in the New York metro area. 
Only students who declared a STEM major at any point in college were included. 

Student data included demographic characteristics, financial status, and academic information such as credits, GPA, major, and grades in gateway science and math courses.  

The key measure of success used was bachelors degree completion, enhanced with information on the time to completion (4 years, 4-6 years, or  $>6$ years), whether the degree was in STEM or not, and whether it was from the original institution or another one.

Our aim was to identify features related to degree completion that may be amenable to improvement through efforts at the institutional level. 

For example, by analyzing the sequences of key gateway courses in STEM and their correlation to degree completion, insights into how institutions might optimize their course sequences and support services could be gained.

Actions based on these insights could improve students’ preparedness for advanced coursework and increase their likelihood of earning their STEM degree.

The approach used in this study to identify factors associated with successful graduation outcomes is a non-statistical ranking method of analysis called $D$-basis, which looks for the strength of associations between input factors (``attributes") and outcome factors (``targets"). 

In previous work, this universal method successfully analyzed subjects described by up to 500 binary attributes \cite{nation2021combining}. 
The advantage of this method is that, unlike traditional statistical methods, large-scale and multifactorial potential factors can be examined simultaneously \cite{adaricheva2017discovery,adaricheva2015measuring}.

Because of the many students who started at the institutions later transferred, data from the National Student Clearinghouse (NSC; www.studentclearinghouse.org) was used to track students' graduation from other institutions. 
NSC data enabled the generation of an accurate and comprehensive view of student success and thereby contributed to informed conclusions about factors influencing students' STEM degree completion.

In this paper, first the data used and our methods of analysis are described.

The results begin with an overview of the various pathways taken by STEM majors in the institutions analyzed.
Then the results of the $D$-basis analysis are described when comparing the outcomes of graduation at any time vs.\ no graduation. %\lh{I took out a mention of 4 years vs. longer here}
An in-depth analysis of factors identified as important follows, including change of majors, grades in key biology, chemistry, and math classes, which math class was taken first, and demographic factors.

Throughout, the implications of these findings for possible interventions that could improve student outcomes are discussed.

This research was carried out collaboratively by faculty from mathematics, computer science, chemistry, and biology departments at higher education institutions on Long Island that are partners in the (STEM)\textsuperscript{2} Network, a National Science Foundation-supported Research Collaboration Network \cite{santangelo2021stem}.

\section{Methods}
\subsection{Data description and collection}

The data analyzed here were obtained from the institutional research offices at the relevant institutions, following IRB approval of our project (Hofstra University IRB 20210126-BIO-HCL-SAN-1).
Data were obtained for first-time full-time students (``FT" students) at two private four-year institutions in the northeastern United States (``School A" and ``School B"), who entered between fall 2010 and fall 2014 and who had either an initial or last major in STEM.  Both institutions are midsized (5,000-15,000 students), selective (undergraduate admission rate 70\%-80\%), and offer degrees in liberal arts and sciences and in professions. 
STEM majors were defined as follows: biological sciences (except medicine and other clinical fields); physical sciences (including physics, chemistry, astronomy, and materials science); mathematical sciences; computer and information sciences; geosciences; engineering; and technology fields associated with the above fields (e.g., biotechnology, chemical technology, engineering technology, information technology).
Because of considerable differences in the patterns of graduation between FT and transfer students, we focused here only on FT students and will present our analysis of transfer student characteristics elsewhere.
Data obtained for each student included general demographic characteristics (gender, race/ethnicity, and eligibility for Pell grants), overall performance attributes (GPA at the end of the first and second semesters, final GPA, retention from year 1 to 2, number of credits attempted and completed, and degree and year completed), and key STEM class information (the first math class taken and grades in the first math class, introductory biology 1 and 2, general chemistry 1 and 2, and organic chemistry).
For students who did not graduate from their home/original institution, students' degree, date of graduation, and major field of degree were obtained from the National Student Clearinghouse, when available.
Our data do not enable us to track students who start with a non-STEM major and leave their original institution, then switch to a STEM major but never graduate.

\subsection{Analysis}

In this work we used an approach to data analysis that involves the discovery of association rules between attributes in binary data.
The advantage of this approach is the possibility of analyzing many attributes that may relate to the outcomes of interest in a study. 
By comparison, in statistical analysis typically only a few attributes can be analyzed together with respect to their correlation with the outcomes.

Association rules became a leading tool of data mining analysis of transaction data
after the introduction of the \emph{Apriori} algorithm 
%(Agrawal, Imielinski, and Swami 
\cite{Agra93}.
Recently Apriori has been included in libraries of R and Microsoft Office.

In our work we employ a novel algorithm called $D$-basis based on
Formal Concept Analysis (FCA) \cite{Wil99}).
This algorithm allows the extraction from a binary table of a set of association rules of confidence = 1 called \emph{implications}. 
We apply frequency analysis to sets of implications to rank attributes of the table in relation to a chosen target attribute: this is a unique feature not available with Apriori.

The $D$-basis algorithm identifies how frequently particular attributes occur in students who had a particular outcome ($d$, the ``target") vs.\ how frequently these same attributes occur in students who had an opposite outcome ($\neg d$, the ``counter-target"). 
For example, the most general test with the data had the target $d=$``cumulative graduation" and the counter-target $\neg d=$``never graduated". The ratio of these two frequencies we call the ``relevance" of a particular attribute with respect to the target and counter-target outcomes.
This ratio is compared to a ``relevance threshold", defined as the ratio of the number of students with the target attribute present to the number of students with the target attribute absent. 
Relevance values above the relevance threshold are considered to reflect an association between the attribute and the target attribute that is stronger than predicted by chance. 
A more detailed description of the $D$-basis method is given in Additional files. 
In the next processing step, the attributes were ordered according to their relevance.

\section{Results and Discussion}
\subsection{Pathways of first-time full-time student retention, majors, and outcomes}\label{aluvial}
The students in our dataset took many paths through higher education, beginning with their initial majors in their original institutions through diverse eventual outcomes.
We defined successful outcomes as those in which students received bachelors degrees at any institution, which National Student Clearinghouse data enabled us to track for the many students who pursued degrees elsewhere after leaving their initial institution.
  \begin{table}[!]
\centering
\renewcommand{\arraystretch}{1.2}
\begin{tabular}{lcc}
   \toprule
   Student Groups & School A & School B \\
   \midrule
   Cohort of 2010-2014 STEM FT students & 1,820 & 1,023  \\
   Not retained after one year & 391 & 165 \\
   Not retained (\% of cohort) & 21.5\% & 16\% \\
   \bottomrule
\end{tabular}
\caption{First-year retention rates among FT STEM students in Schools A and B.}
\label{tab:Retain}
\end{table}

\begin{figure}[!]
    \centering
    \includegraphics[width=\textwidth]{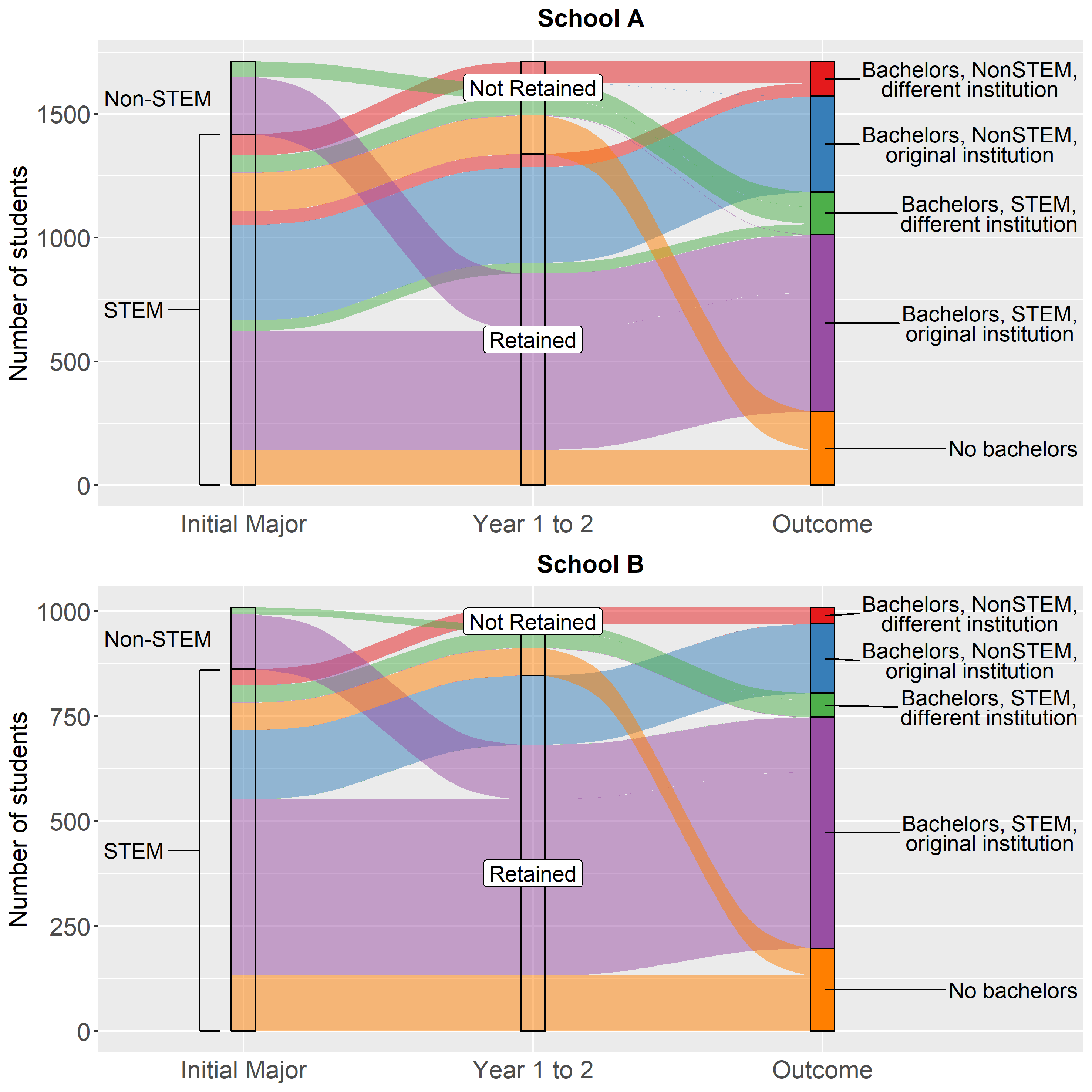}
    \caption{Majors, retention, and graduation outcomes of first-time full-time (FT) students at Schools A and B. 
    The left  bar represents all the FT students at each school in the 2010-2014 starting cohorts who either started as STEM majors or as non-STEM majors who eventually switched to a STEM major (non-STEM majors who never switched to a STEM major are not included).
    The center bar indicates whether students were retained from the first to the second year.
    The right bar represents the eventual outcomes.
    Colored streams represent groups of students with shared characteristics.} 
    \label{fig:AlluvialBothSchoolsV3}
\end{figure}

Students' paths from entry through the transition from the first year to the second year and then to the last available undergraduate outcome were visualized using alluvial plots (Figure \ref{fig:AlluvialBothSchoolsV3}). 
After one year, the majority of each cohort was retained at the original institution but 21.5\% (School A) or 16\% (School B) were not retained (Table \ref{tab:Retain}). 
Almost half of all students in the group of students not retained after one year were those who started as STEM majors and who ended up not transferring to other colleges and never graduating. 
At School A they make up about half of those who stayed with STEM and never graduated. 
At School B the proportion of STEM majors not retained after one year is smaller, and these non-retained students make up a smaller percentage of all non-graduates compared to School A.
These students who are not retained at their original institution after one year show a variety of outcomes, which makes the analysis of retention a difficult task.  
We plan to look more closely at this in a future publication.
 The two orange streams on the alluvial plots represent the largest loss of STEM majors in these two institutions 

Most of the students who graduated in STEM had also chosen to major in STEM initially, but close to one-fifth of these STEM graduates 
had an initial non-STEM major (which included those who were undecided; purple streams in the plots). 

The blue stream represents initial STEM majors who switched to non-STEM by graduation at their original institution. 
This group of students was larger at School A than at School B. 
At both schools these groups are much larger than the groups of students who switched from non-STEM (mostly undecided) to STEM majors.
Detailed analysis (to be described in section \ref{sec:sw of Mj}) showed that switching major was strongly associated with graduation outcomes.

The green streams represent non-retained students who successfully graduated at other institutions, either staying in STEM or switching from non-STEM into STEM. 
One subgroup of them was retained after one year, which indicates that many still transferred after the second year from both schools.
Finally, the red stream represents non-retained students who switched their major and graduated as non-STEM majors. 
Again, a smaller portion of them were retained after one year.

\subsection{\textit{D}-basis ranking of attributes associated with educational outcomes} \label{D-basis analysis}

 \subsubsection{Targeted outcome: graduation at any time versus never graduated}

\begin{figure}[!ht]
    \centering
    \includegraphics[width= \textwidth]{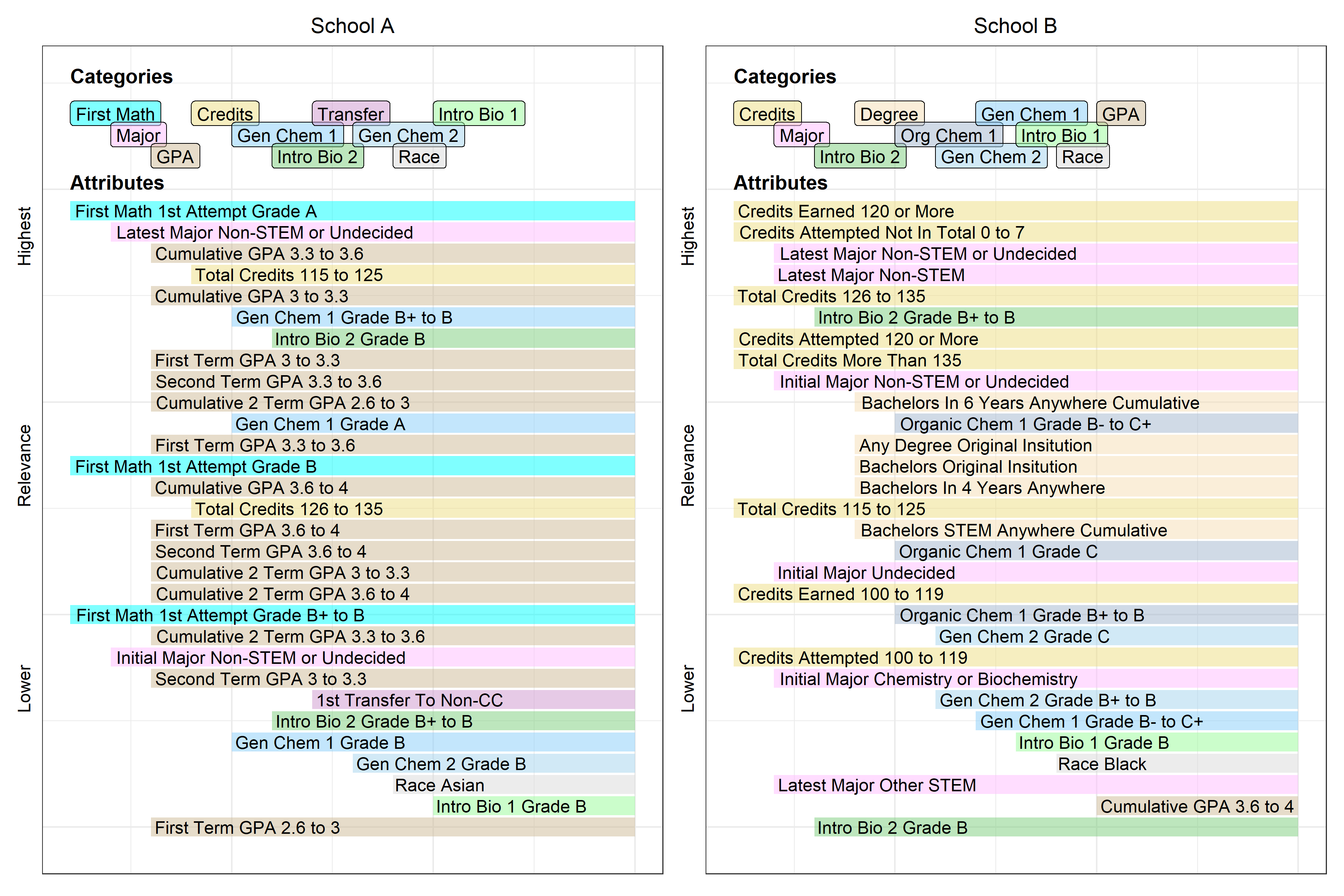}
    \caption{Top 30 attributes (out of 206) for graduation vs.\ non-graduation in order of decreasing relevance by $D$-basis analysis. Categories of related attributes are represented in the header as rectangles. The length and color of the attribute bars correspond to category of the attribute and match the category rectangles at the top (e.g., all First Math-related attribute bars for School A are colored in turquoise and shown at far left aligned with the First Math category rectangle). For full list, see (
    \url{https://tinyurl.com/2j3hx4yz/SchoolA\_FTFT\_STEM-ByRelevance\_col11-col13-FullNames.csv}
    and 
    \url{https://tinyurl.com/2j3hx4yz/SchoolB\_FTFT\_STEM-ByRelevance\_col11-col13-FullNames.csv})}
    
    \label{fig:RankBothSchoolsV2}

\end{figure}

 Because many attributes in the original data are split into multiple attributes when converted to binary data for use in the analysis (e.g., the grade in a specific course could be split into grade of A, grade of B, etc.), post-$D$-basis clustering of related attributes in the attribute ranking was carried out to aid in interpretation. 
 An example output for Schools A and B, with related attributes grouped by color and indent and the grouping categories shown in rectangles at the top, is shown in Figure \ref{fig:RankBothSchoolsV2} (top 30 shown out of 206 tested). 
 Some highly-ranked attributes are expected, such as high numbers of total credits.

 For School A FT students who started or finished with a STEM major, when targeting $d=$``cumulative graduation" and $\neg d=$``not graduated", the following attributes were highly associated with graduation (Figure \ref{fig:RankBothSchoolsV2} and additional results not shown):

\begin{itemize}
     \item Good grades in the first mathematics course, in introductory biology 1 and 2, and in general chemistry 1 and 2 (introductory college chemistry). 
     Indeed, the highest ranked attribute of all for these students was earning an ``A" in the first attempt at the first math class. 
     Note that there was a high percentage of chemistry and biology majors present in the data set.
     \item Finishing or beginning with a non-STEM major. 
     As all these students were STEM majors at some point, the students who ended with a non-STEM major must have changed from an initial STEM major.
\item High GPAs (3.0 or above) during the first two semesters.
\item Transferring to a non-community college: these students transferred out of their original institution and graduated elsewhere.
\item Identifying as Asian
\item Not in the top 30 attributes, but also ranking highly, were the selection of the first math class: taking calculus II as the first math class, followed by calculus III, then calculus I, and precalculus (not shown in figure).
 \end{itemize}

 For the same School A FT STEM major students, when targeting $d=$``not graduated" and $\neg d=$``cumulative graduation", the following attributes were highly associated with not graduating (data not shown):
\begin{itemize}
     \item Not retained between year 1 to 2;
     \item Credits attempted up to 39;
     \item Low GPA the first 2 semesters;
     \item Not transferring to another college;
     \item Failing grades in general chemistry 1, introductory biology 2, or the first math class;
     \item A high rate of attempting and not completing courses.
 \end{itemize}

For School B's FT STEM majors, the attributes most associated with graduation were as follows (Figure \ref{fig:RankBothSchoolsV2}): 

\begin{itemize}
    \item Finishing or starting with a non-STEM major
    \item Good grades in organic chemistry, general chemistry, and biology. 
    \item Good GPA overall and in the first two semesters
    \item Identifying as Black (note that the number of Black students in this population was small) 
\item Not being eligible for Pell grants (i.e., low financial need; data not shown)% in Figure \ref{fig:RankBothSchoolsV2})
\end{itemize}

When targeting $d=$``not graduated" and $\neg d=$``cumulative graduation for School B students, highly associated attributes included the following (data not shown):
\begin{itemize}
    \item Low GPA first two semesters
    \item Poor grades in introductory biology, general chemistry, and the first attempt of the first math class 
    \item Not transferring to another college and transferring to a community college
\end{itemize}

Both similarities and differences were observed when comparing the attributes associated with graduation in School A vs.\ School B. 
The grade in the first math class and choice of the first math class are very important for graduation in School A but not in School B, whereas the grade in organic chemistry is much more important in School B than in School A. 
General chemistry and introductory biology grades are important in both schools, as are credits completed and GPA during the first two semesters and overall. 
An unexpected and important finding for these STEM majors at both schools is that starting or finishing with a non-STEM major (including undecided) is highly associated with graduation. 
We examine some of these findings in more detail in Section \ref{Analysis of important attributes}.

We carried out a similar D-basis analysis with target $d=$``graduation in 4 years" and $\neg d=$``graduation in more than 4 years".
Results of this analysis are reported in the Additional files. 

 \vspace{1cm}
\FloatBarrier
\subsection{Detailed analysis of academic attributes associated with graduation}\label{Analysis of important attributes}

 \subsubsection{Switch of major}\label{sec:sw of Mj}
 The $D$-basis analysis showed that the attribute of switching from an initial STEM major to a non-STEM major was highly relevant to graduation for both schools (Figure \ref {fig:RankBothSchoolsV2}).  
 An initial major of Undecided also appeared quite high in the ranking by relevance when targeting graduation. 
Following up on this $D$-basis result,
 we examined the outcomes for two subgroups more closely: those whose initial majors were STEM and those who were initially undecided and later chose a STEM major. The sizes of these groups are shown in Table \ref{tab:Groups}.

\begin{table}[h!]
\centering
\renewcommand{\arraystretch}{1.2}
\begin{tabular}{lcc}
   \toprule
   Student Groups & School A & School B \\
   \midrule
   Cohorts of 2010-2014 (number of students) & 10,853 & 4,734  \\
   Initial major STEM (\% of cohort) & 18\% & 18\% \\
   Initial major Undecided (\% of cohort) & 18\% & 24\% \\
   \addlinespace
   Latest major STEM (\% of Initial STEM) & 65\% & 80\% \\
   Latest major STEM (\% of Initial Undecided) & 9\% & 9\% \\
   Latest major STEM (\% of the whole cohort) & 16\%  & 18\% \\
   \bottomrule
\end{tabular}
\caption{Subgroups of Initial STEM majors and Undecided in Schools A and B.\\ ``Initial Undecided" includes all undecided students regardless of eventual major.}
\label{tab:Groups}
\end{table}

 \begin{table}[h!]
\centering
\renewcommand{\arraystretch}{1.2}
\begin{tabular}{l ccc ccc}
   \toprule
   & \multicolumn{6}{c}{Time to graduation} \\
   \cmidrule(lr){2-7}
   & \multicolumn{3}{c}{School A} & \multicolumn{3}{c}{School B} \\
   \cmidrule(lr){2-4} \cmidrule(lr){5-7}
   Initial and Latest Majors & 4 yr & 6 yr & $>$ 6 yr & 4 yr & 6 yr & $>$ 6 yr \\
   \midrule
   Initial STEM + Latest STEM & 47\% & 64\% & 69\% & 51\% & 60\% & 61\% \\
   Initial STEM + Latest non-STEM & 68\% & 86\% & 89\% & 58\% & 74\% & 75\% \\
   \addlinespace
   Initial Undecided + Latest STEM & 54\% & 81\% & 87\% & 61\% & 90\% & 91\% \\
   Initial Undecided + Latest non-STEM & 59\% & 75\% & 79\% & 44\% & 59\% & 61\% \\
   \bottomrule
\end{tabular}
\caption{Cumulative graduation rates as a proportion of the students in each subgroup.} 
\label{tab:GradABcombined}
\end{table}

Analysis of graduation rates for student with initial STEM majors shows that those who switched their major to non-STEM had much higher graduation rates than those who stayed in STEM majors: a 20-22\% gap for School A and a 7-14\% gap for School B (Table \ref{tab:GradABcombined}; compare 1st and 2nd rows). 
The opposite result was obtained for those students who were initially undecided as to major: those who had a latest major in STEM had higher graduation rates compared to those whose latest major was non-STEM, except for School A's 4 year graduation rate (54\% vs.\ 59\%, Table \ref{tab:GradABcombined}; compare 3rd and 4th rows in 1st data column). 
For School B this gap was 17-31 points (Table \ref{tab:GradABcombined}; compare 3rd and 4th rows, 4th-6th data columns).

We hypothesize that some students feel pressure to enter college with a STEM major and to stay in these majors despite performing poorly in STEM classes, leading them to eventually drop out of college and never graduate \cite{Seymour_2019}, \cite{Pitt_2021}. 
Encouraging these students to switch to other majors may enable them to graduate successfully, even though in different disciplines.
The higher rates of success in STEM for those who enter as undecided and choose STEM at later times suggest that universities should be careful when advising these students and not pressure them to declare a major until the students have tried introductory courses in their prospective STEM major. 
If they are not successful in these courses they might choose alternative majors in which their likelihood of success is higher.

\subsubsection{Grades in biology and chemistry classes}

\begin{figure}[h!]
\begin{center}
     \includegraphics[width=\textwidth, height=10 cm]{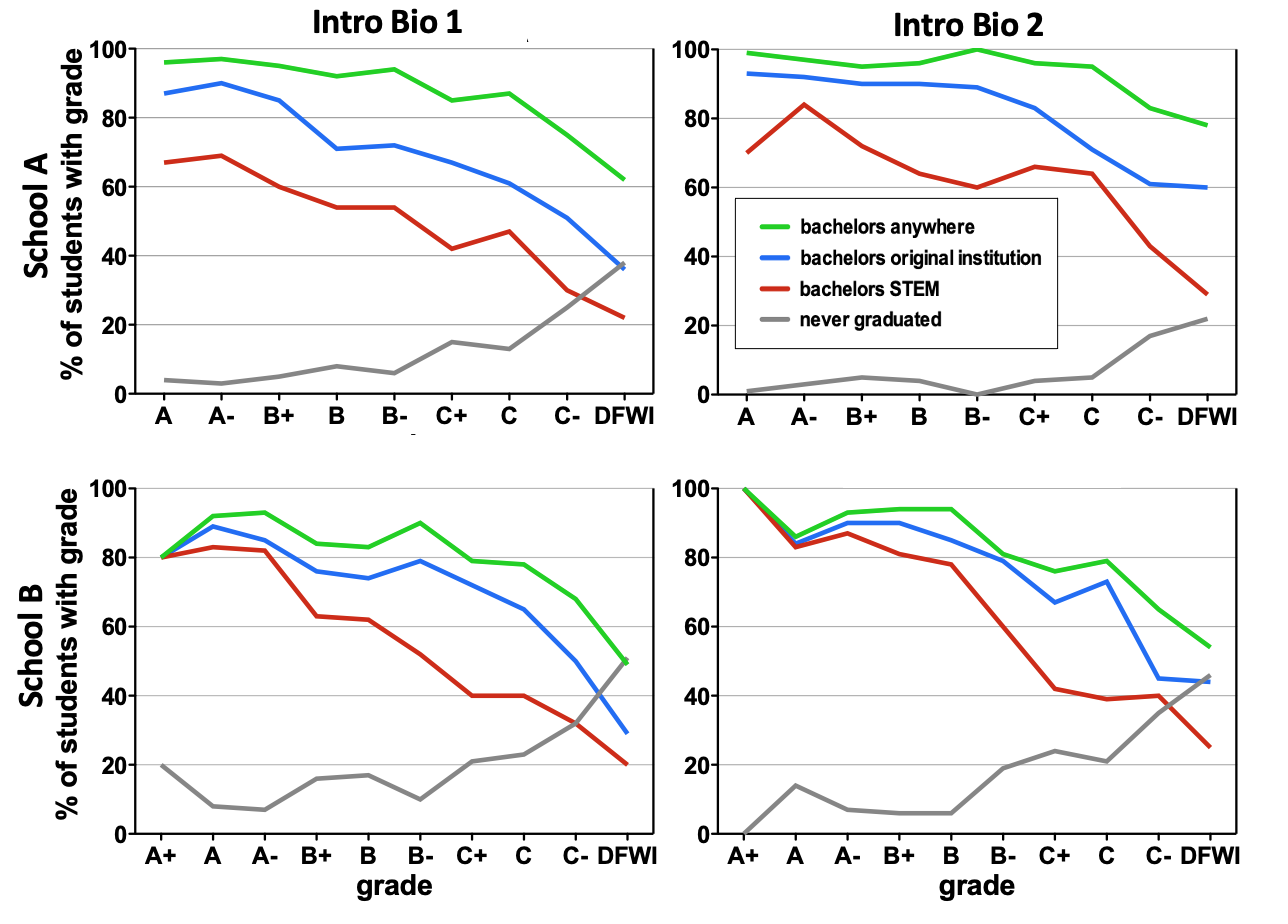}
\end{center}
\caption{Introductory Biology 1 and 2 course grades at Schools A and B and student outcomes. ``DFWI" means grades of D, F, withdrew (W), or incomplete (I).}
\label{fig:June24IntroBio}
\end{figure}

\begin{figure}[!]
\begin{center}      \includegraphics[width=\textwidth, height=8 cm]{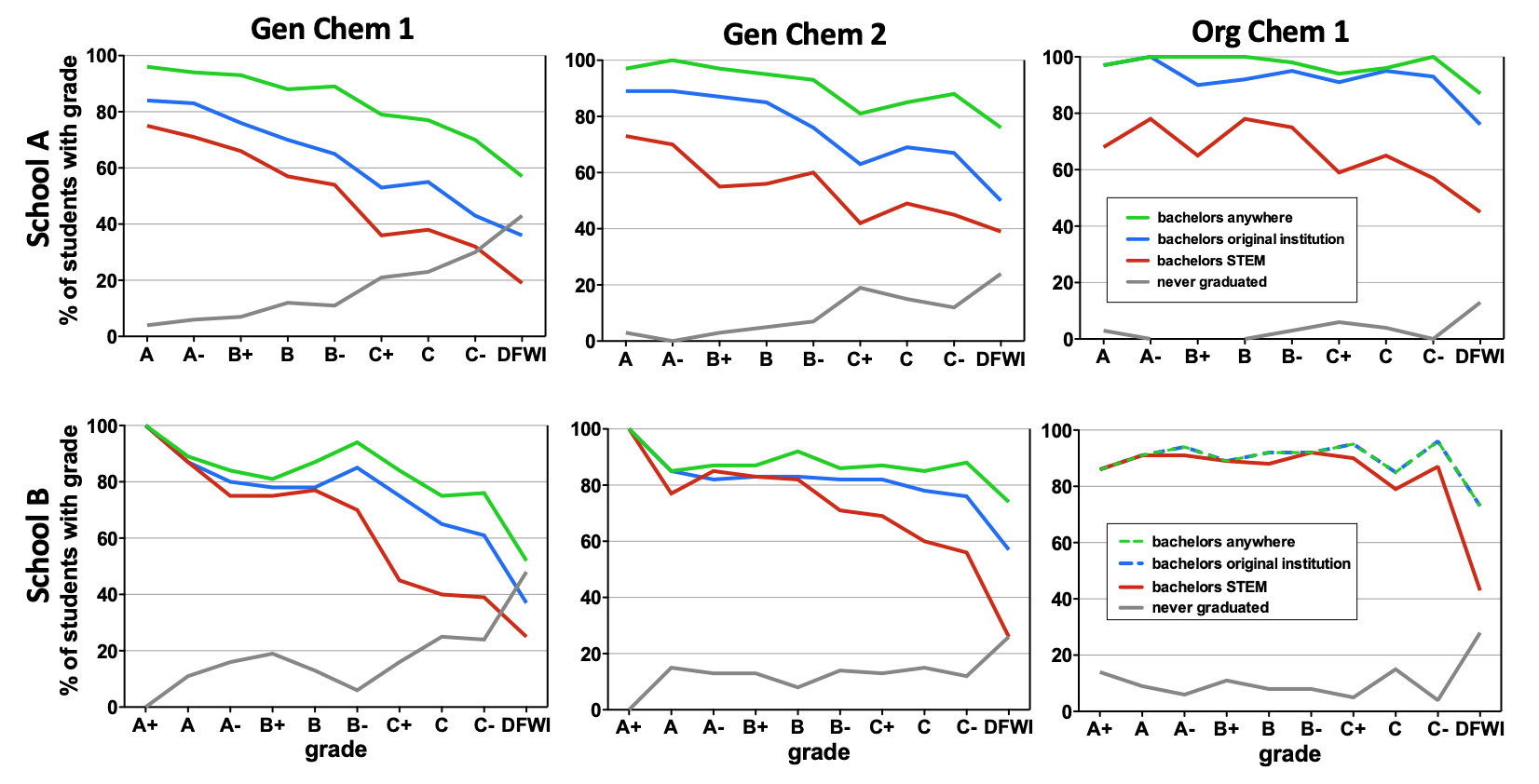}
\end{center}
\caption{General chemistry 1, 2 and organic chemistry course grades and outcomes at Schools A and B. School B org chem 1, blue and green lines are depicted as dotted lines as the values are identical. ``DFWI" means grades of D, F, withdrew (W), or incomplete (I).}
\label{fig:June24Chem}

\end{figure}

The $D$-basis results showed that grades in five ``gateway" STEM classes--introductory biology 1 and 2, general chemistry 1 and 2, and organic chemistry 1--ranked highly in their association with graduation in both schools (Figure \ref{fig:RankBothSchoolsV2}).
Therefore, we analyzed in more detail the outcomes for School A and School B STEM majors as a function of the grades these students obtained in these classes. 
The four possible outcomes examined were bachelors (graduation) anywhere, bachelors from the original institution, bachelors in STEM, and never graduated (Figures \ref{fig:June24IntroBio}, \ref{fig:June24Chem}).

To summarize these results in general: grades in the first semesters of introductory biology and general chemistry showed strong correlation to student graduation outcomes at both institutions, 
but the details of these relationships differed by course and by institution.
 The most sensitive outcome, i.e., the first to drop as grades in these introductory courses decreased, was graduation with a STEM degree, followed by graduation from the original institution.
 Graduation with a bachelors anywhere was the least sensitive to grades in these introductory courses. 
 Grades in the second semesters of introductory biology and general chemistry also showed correlations with successful outcomes but generally the rates of successful outcomes were higher than for the same grades in the first semester (in Figure \ref{fig:June24IntroBio}, compare School A, introductory biology 1 and 2, and in Figure \ref{fig:June24Chem} compare general chemistry 1 and 2 at both schools).
 We suggest that this is likely because in order to take the second semester of introductory biology or general chemistry, students must usually have obtained a minimum passing grade (e.g., C-) in the first semester, thus already demonstrating the ability to succeed in a college STEM class.
 The rosters of the second semester courses thus lack many of the students who were least likely to be able to successfully complete a STEM bachelors. 

 Grades of C- and above in organic chemistry 1 generally correlated with success in graduating anywhere.
 This observation may appear surprising given the formidable reputation of organic chemistry 1 as a challenging course. 
 We speculate that students who made it to organic chemistry 1 had already demonstrated significant academic ability and almost all who passed it would therefore be able to earn a bachelors degree.
 
 Finally, the rate of ``never graduated" generally rose as grades decreased, with the most striking increases in ``never graduated" occurring for grades of DFWI in all courses.
 In particular, students who received DFWI grades in introductory biology 1 or general chemistry 1 at both schools had close to or greater than 40\% rates of never graduating, suggesting that improving student success in these courses could result in increases in successful long-term outcomes for students.

\subsubsection{First mathematics class}

Numerous studies have demonstrated the importance of students' mathematics preparation and aptitude for their success in college STEM courses \cite{paschal2021examination}, \cite{crisp2009student}, \cite{Wu2023},  \cite{Spencer1996},  \cite{Salehi2019}.
 
The level of preparation is reflected in which mathematics class students take in their first year.

At both schools biology majors have fewer math requirements than other STEM majors. 
At School A, biology majors must take three math-related courses: biostatistics (offered by Biology), precalculus or calculus, and one additional mathematics class. 
At School B, biology majors must take two math-related courses: elementary statistics (offered by Mathematics) and either precalculus or calculus (or, rarely, computer science). 
Given that biology majors make up a large proportion of STEM majors in both schools, the less demanding math requirements at School B likely explain why at School B math course choice and grades were not identified as highly ranking factors associated with graduation  (Figure \ref{fig:RankBothSchoolsV2}). 

Most STEM majors other than biology require at least calculus II, which is either a terminal mathematics class or a prerequisite for other classes for majors in mathematics, physics and engineering. 
For students in these majors, successful progression in the major therefore generally means passing calculus I in their first semester. 

We first examined the association of which math class a student took first (the ``choice" of first math class) 
with graduation outcomes both in the $D$-basis results and through analysis of different groups of majors.
The choice of a student's first math class is influenced and often constrained by many factors including the student's preparation, performance on a placement exam (required at both schools), requirements of the planned major, and advising.  
We present the results for School A and School B separately because of the significant differences between the schools in their math requirements and in the results found in our analysis.
We then examine the association of grades in the first math class with graduation outcomes.
\subsubsection{Choice of first math class at School A}\label{firstmathchoiceA} 
In the analysis of graduation vs.\ non-graduation, 
the first math class attributes in School A with rankings above the threshold were as follows 
(the rank appears in parentheses next to the attribute, with a higher rank corresponding to a smaller number):\\

For all STEM majors:\\
Calculus II (46) $>$ Calculus III (78) $>$  Calculus I (93) $>$ Precalculus (99) $>$ Statistics (119).\\

For biology majors only:\\
Calculus I (6)$>$ Gen Ed math (48)$>$ Calculus II (72) $>$ Precalculus (89) $>$ Calculus III (92).\\

These results show that for biology majors at School A, taking calculus I as the first math class is highly associated with graduation, whereas for all STEM majors, taking calculus II or III as the first math is more highly associated with graduation than is calculus I. 
Furthermore, precalculus as a first math class ranks quite low in its association with graduation compared to other mathematics courses for both biology and all STEM majors.

Consistent with this $D$-basis result, the graduation rates of STEM majors, both biology and non-biology, dropped when their first math was precalculus compared to any calculus courses 
(Table  \ref{TabSTEMcombinedMathA}).

\begin{table}[t]
\centering
\renewcommand{\arraystretch}{1.2}
\begin{tabular}{l cccc cccc}
   \toprule
     & \multicolumn{4}{c}{STEM non-Bio majors} & \multicolumn{4}{c}{Bio majors} \\
     \cmidrule(lr){2-5}  \cmidrule(lr){6-9}
     First math class & Students & 4 yr & 6 yr & $>$ 6 yr & Students & 4 yr & 6 yr & $>$ 6 yr \\
   \midrule
   All math classes & 1,018 & 44\% & 59\% & 60\% & 803 & 56\% & 78\% & 82\% \\
   Calculus I,II, or III & 557 & 57\% & 81\% & 84\% & 245 & 67\% & 85\% & 89\% \\
   Precalculus & 219 & 42\% & 70\% & 78\% & 233 & 53\% & 78\% & 81\% \\
   Other math class & 183 & 44\% & 71\% & 79\% & 169 & 49\% & 73\% & 79\% \\
   No math class & 60 & 25\% & 25\% & 25\% & 155 & 43\% & 45\% & 45\% \\
   \bottomrule
\end{tabular}
\caption{Graduation rates of School A STEM non-Bio and Bio majors according to their\\ first mathematics class, as a percentage of each group.}
\label{TabSTEMcombinedMathA}
\end{table}

The lowest graduation rates of all among STEM majors were within the subgroup of 60 non-biology majors and 155 biology majors who did not take any math class (Table  \ref{TabSTEMcombinedMathA}).
Only about 25\% of non-biology majors and 45\% of biology majors in this subgroup graduated with a bachelor's degree, 
but less than 10\% overall did so at their original institution, and only one student in a STEM discipline (data not shown). 
Many in this subgroup were not retained after the first year, partly explaining why there was no record of a math class. 

To summarize the observations on the choice of first math class for School A STEM students: students across all majors had higher graduation rates when they started their math education at calculus I or higher compared to when their first math class was precalculus, other mathematics classes, or no math at all. 

\subsubsection{Choice of first math class at School B}\label{firstmathchoiceB}

The $D$-basis analysis for School B using graduation  vs.\ non-graduation as outcomes gave the following relevance ranks for choice of first math class:\\ 

For all STEM majors:\\
Calculus II (44) $>$ Advanced math (53) $>$ Calculus III (68) $>$ Statistics (70) $>$ Calculus I (92) $>$ Precalculus (93)\\

For biology majors:\\
Calculus I (40) $>$ Statistics (71) $>$ Precalculus (77) $>$ Calculus II (91)\\

For all STEM majors, calculus II and III and advanced math classes are the most highly associated with graduation at any time, whereas for biology majors calculus I followed by statistics are most highly associated. 

In School B, statistics is taught by the Mathematics Department and required for biology and several other majors. 
Statistics was therefore the most popular first math course for biology majors and was also the first math class for 97/486 (about 20\%) of STEM non-biology majors. 
Especially for these students, Statistics was associated with high graduation rates at 4, 6 and $>6$ yr (Table \ref{TabSTEMcombinedMathB}).

\begin{table}[h!]
\centering
\renewcommand{\arraystretch}{1.2}
\begin{tabular}{l cccc cccc}
   \toprule
   & \multicolumn{4}{c}{STEM non-Bio majors} & \multicolumn{4}{c}{Bio majors} \\
   \cmidrule(lr){2-5} \cmidrule(lr){6-9}
   & & \multicolumn{3}{c}{Time to graduation} & & \multicolumn{3}{c}{Time to graduation} \\
   \cmidrule(lr){3-5} \cmidrule(lr){7-9}
  First math class & Students & 4 yr & 6 yr & $>$ 6 yr & Students & 4 yr & 6 yr & $>$ 6 yr \\
   \midrule
   All math classes & 486 & 60\% & 78\% & 80\% & 537 & 60\% & 76\% & 79\% \\
   Calculus I,II, or III & 290 & 58\% & 76\% & 78\% & 54 & 63\% & 81\% & 83\% \\
   Precalculus & 26 & 54\% & 69\% & 73\% & 93 & 55\% & 75\% & 77\% \\
   Statistics & 97 & 69\% & 92\% & 93\% & 293 & 67\% & 80\% & 81\% \\
   Other math class & 50 & 76\% & 90\% & 90\% & 13 & 54\% & 85\% & 92\% \\
   No math class & 23 & 26\% & 35\% & 48\% & 84 & 42\% & 57\% & 67\% \\
   \bottomrule
\end{tabular}
\caption{Graduation rates of School B STEM non-Bio and Bio majors according to their\\ first mathematics class, as \% of each group} 
\label{TabSTEMcombinedMathB}
\end{table}

The graduation rates were very similar between biology majors who took either statistics or calculus, but those who took precalculus had lower graduation rates, as seen also at School A.
Comparing Tables \ref{TabSTEMcombinedMathA}  and \ref{TabSTEMcombinedMathB} for biology majors of the two schools, only 27\% of biology majors take precalculus/calculus class as their first class in School B, while 60\% do in School A, a difference that may contribute to higher graduation rates at School B.

\begin{center}
\begin{figure}
      \includegraphics[width=\textwidth, height=10 cm]{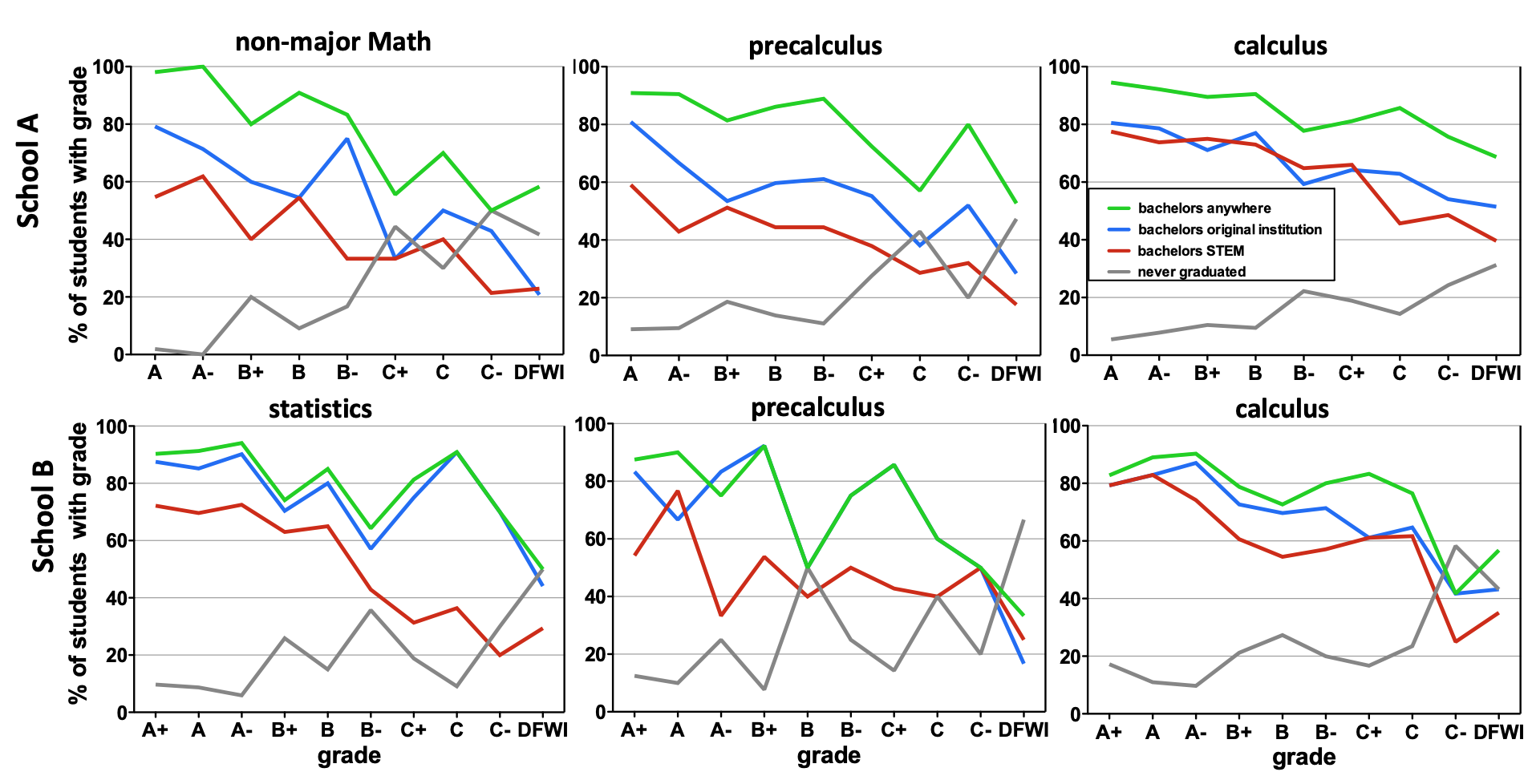}
      
\caption{First math course grades at schools A and B and outcomes. Trend lines of \% students with specific grades with outcomes. 
}
\label{fig:mathgraph}
\end{figure}
\end{center}

\subsubsection{Grades in first math class}

The same type of analysis which was done for the grades in biology and chemistry courses was done for mathematics classes. 
For the math classes, however, we only have the grade in the first math class. 
We grouped the first math classes into three categories for each school.

For School A, the categories are non-major mathematics, precalculus, and calculus (any level).  
For School B, the categories are statistics, precalculus, and calculus (any level). 
Non-major math at School A was replaced in our categorization by statistics at School B because relatively few students  took statistics at School A or a non-major mathematics class at School B (Figure \ref{fig:mathgraph}).

For both schools, students who took calculus as the first math and earned grades of C or higher had high rates of completing a bachelors degree (Figure \ref{fig:mathgraph}). 
This observation is consistent with results of sections \ref{firstmathchoiceA}  and \ref{firstmathchoiceB} that taking a calculus class as the first math class
relates to the highest graduation rates.

For School A, even students who earned  C- or DFWI in calculus as their first math class mostly received a bachelors degree.  
At School B, however, there was a sharp drop in positive outcomes below a grade of C.

The frequency of earning a STEM bachelors degree showed different patterns between the two schools according to the grade in calculus.  
At School A this rate declined below a grade of B, with a steep drop between C+ and C.  
However, even for DFWI grades, 40\% of students went on to earn a bachelors in STEM somewhere.  
At School B, grades of A- and below in calculus were associated with decreases in the rate of earning a STEM bachelors, with the sharpest drop between those earning a C and C-.

Looking across each row,  at School A in particular, there is a larger gap between earning a bachelors anywhere (Figure \ref{fig:mathgraph}, top row, green line) and earning a bachelors in STEM (red line) for grades in the first math class for non-major mathematics classes or precalculus than for calculus. 
School B results were similar although with more variability across grades.
A student's grade in a first non-calculus mathematics class appears much more related to overall success (graduation) than to graduation with a STEM major. 

This finding is consistent with  some of the students taking such a class switching their major from STEM to non-STEM, and in section \ref{sec:sw of Mj} we found that switching majors is associated with higher graduation rates.
A closer examination confirmed that in School A more students taking precalculus or general education math classes switched their major than stayed in their major, and the graduation rate in the first group was  75\%, while in the second group it was 24-40\%.  In School B fewer than half the students taking precalculus or statistics switched their major,  but those that switched had a higher graduation rate, approximately 90\%, than those that did not, 55-71\%.

For students at School B whose first math was statistics, grades below C led to a reduced rate of earning a bachelors degree and grades below a B led to a sharp drop in the percentage of students earning a bachelors in STEM, suggesting that a student's grade in statistics can indicate their chance of successful outcomes overall.
This is in accordance with analysis in section \ref{firstmathchoiceB}, where we see that taking statistics as the first math class at School B is an important predictor of the graduation outcome.

\begin{figure}[!ht]
    \centering
    \includegraphics[width=\textwidth]{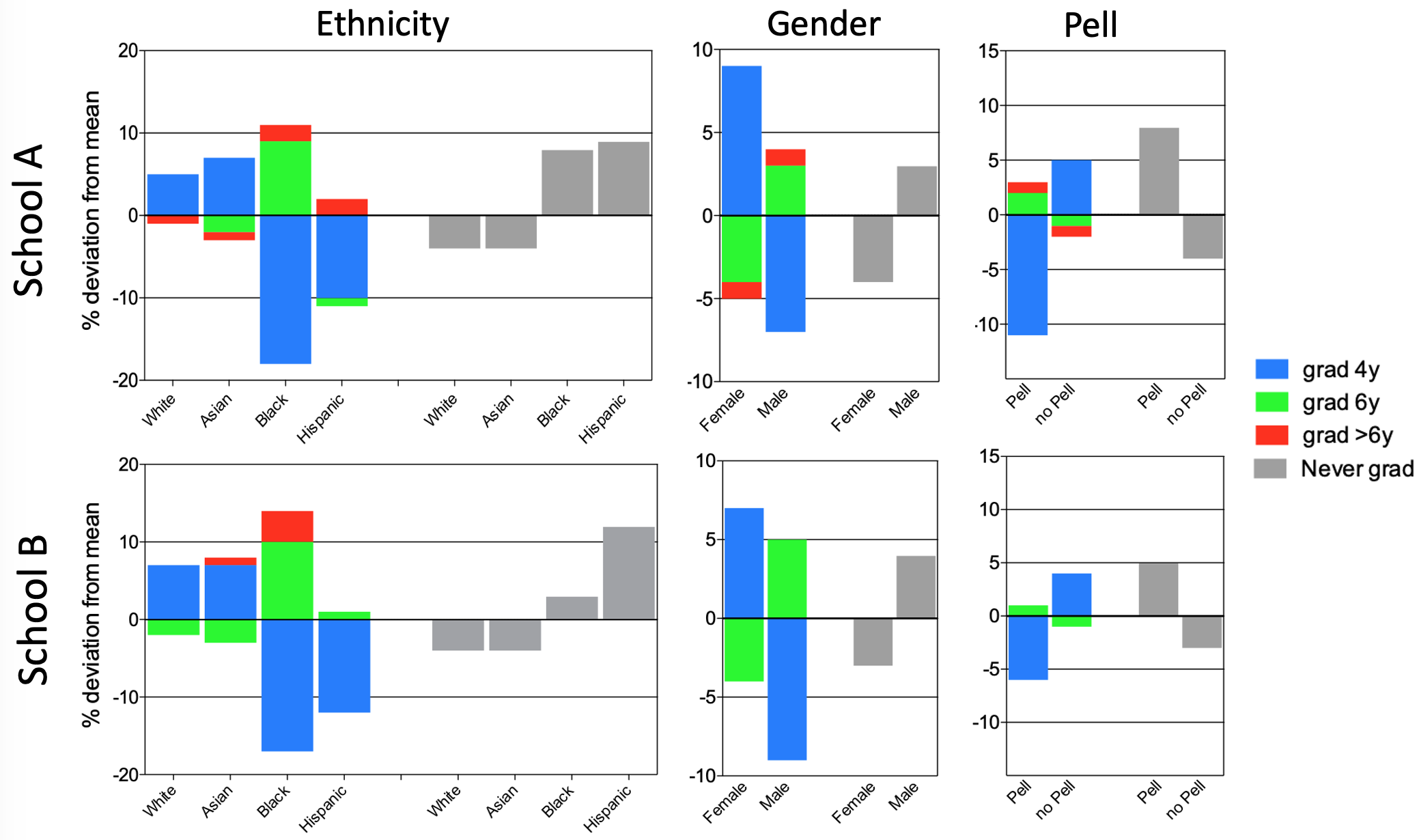}
    \caption{Correlations of demographic factors and outcomes among FT STEM majors.  Percent of deviation from the mean among the same category factors from Schools A and B, respectively.  Racial categories are White, Asian, Black and Hispanic. Gender categories are female (F) and male (M). Pell categories are Pell-eligible (``Pell")  and not-eligible (``no Pell").  The figure legend shows in blue for graduation by 4 years (grad 4y), in green, by $>$4 to 6 years (grad 6 y), in red, graduation after 6 years (grad $>$ 6y), and in grey, never graduated (never grad).
    }
\label{fig:demographics}
\end{figure}

\subsection{Analysis of demographic factors}
Past analyses from many studies indicated that historically disadvantaged minorities, female, first generation, and economically challenged students have more barriers to success compared to students who are White or Asian, male, continuing generation, or economically advantaged \cite{bottia2021factors}, \cite{campbell2022how}, \cite{doi:10.1080/2331186X.2016.1178441}, \cite{https://doi.org/10.1002/tea.21301}, \cite{10.17763/haer.81.2.324m5t1535026g76}.
Surprisingly, $D$-basis ranking analysis that compared bachelor’s degree graduation anywhere inclusive of any time ( $>$ 6 years) to those who never graduated (Figure \ref{fig:RankBothSchoolsV2})
showed that the usual demographic factors were not highly associated with graduation for either school examined. 
The $D$-basis analysis was followed up with detailed analysis of students in different demographic groups and their various outcomes.
The results of this detailed analysis (Figure  \ref{fig:demographics}) are presented as deviations from the mean values for all students, i.e., a positive value indicates that the indicated outcome occurred more often for the specified group of students than average.
The trends and proportions of the analyzed outcomes were very similar between School A and School B and also similar for all STEM majors and biology majors only, so we present only the results for all STEM majors.
\\
\subsubsection{Race and ethnicity}
$D$-basis analysis of racial and ethnic-related attributes for all STEM majors showed that for School A, identifying as Asian was strongly associated with graduation anywhere vs.\ never graduating (rank of 30 among all attributes) whereas for School B, identifying as Black ranked 27 (see Figure \ref{fig:RankBothSchoolsV2}).\
However, there were only 18 Black students in STEM who did not graduate at School B, and we have seen in other analyses that $D$-basis ranks can appear inflated with small numbers in the sample.

Other racial classifications ranked as follows in their association with graduation: 
at School A, White (72), Black (98) and Hispanic (105, close to the threshold), and at School B,
White (63), Asian (65) and Hispanic (75). 

Detailed analysis showed that White and Asian students took a shorter time to graduate than Black and Hispanic students, as seen by the positive values (i.e., values above the mean of all students) for graduation in 4 years for both of these groups of students in Figure  \ref{fig:demographics}.
Among the never graduating groups, Hispanic students had the largest increase above the mean, followed by Black students. 
At School A, Black and Hispanic students never graduated at high and similar rates whereas at School B, the rate at which Hispanic students never graduated was greater than that of Black students. 
\\

\subsubsection{Gender}
For the years studied, the only gender identifications provided by the institutions were male and female, although we recognize that many students' identities do not fall into this binary distinction.
$D$-basis analysis of  4 years vs.\ later graduation found that identifying as female ranked 1 in relevance for School B, whereas male was 83.
In the detailed analysis, the trends were similar at both schools (Figure  \ref{fig:demographics}). 
Female students took less time to graduate than male students (overall and at their original institutions) and more male students never graduated. 
Further analysis indicates that despite higher switching rates out of STEM, female students have an overall higher graduation success rate than males in STEM majors and when they switch from STEM to non-STEM majors.  
\\

\subsubsection{Pell eligibility}

We examined Pell eligibility status to assess the effect that economic factors may have on student outcomes. 
About 30\% of STEM majors at both schools were eligible for Pell grants. 
 Pell eligibility was among the top 30 ranked attributes for graduating within 4 years compared to those that graduated later in our $D$-basis analysis for School B (Figure \ref{fig:RankedAttributesCol6vsCol9WHeaders-7-23-24}), but this association between Pell and earlier graduation seems to be somewhat artificially inflated as a result of the small number of students who were Pell eligible and graduated in more than 4 years at School B.
In the detailed analysis we see that at both schools,
Pell-eligible students took longer to graduate and a larger fraction also never graduated (Figure  \ref{fig:demographics}), similar to others' findings \cite{nichols2015}.
In another report, the effect of a Pell grant on student success was shown to be state  dependent, possibly owing to the interaction between Pell grants and state aid programs \cite{doi:10.1086/712556}.
Other studies suggest that Pell-eligible students are more likely to share other challenges associated with lower degree completion rates \cite{colvard2018}.\\

\section{Conclusions}

In this paper we used an algorithm known as $D$-basis analysis to simultaneously examine multiple attributes describing the cohorts of first time students choosing a STEM major in 2010-2015 in two private 4-year higher education institutions in the northeastern United States. 
This approach allowed us to rank all attributes in relation to target attributes, specifically graduation with a bachelors degree vs. non-graduation.

Most of the attributes included in the study were related to the choice of and performance in several gateway classes in STEM education, as well as success markers of the first year in college such as the number of completed credits, GPA in the first two semesters, and retention from year 1 to 2.
Other attributes described students' educational paths, such as whether they changed their major or transferred to another school and how long they took to graduate and whether they did so from their original institution or not. 
Some demographic factors were also included.

The goal of the study was to reveal attributes that could affect students' likelihood of graduating, specifically those that are amenable to interventions by the schools, and as a result to improve graduation rates among students who choose a STEM major at least for part of their student career. 
We used a broad and inclusive definition of success, namely graduation with a bachelors degree, independent of major and of the final school of graduation. 
  Here we summarize several discoveries revealed by the ranking of attributes through the $D$-basis analysis (subsection \ref{D-basis analysis})
  and detailed examination of the original data (subsection \ref{Analysis of important attributes}).

\begin{itemize}
    \item [(1)] Changing majors is important for some students to have a successful outcome, namely graduation with any bachelors degree \emph{even if not with a STEM degree}. 
    This observation suggests that institutions might promote exploration of majors through course selection and, for students who are not doing well, flexibility in changing majors early.
    \item[(2)] There were two main findings about the first mathematics class and student outcomes: first, STEM majors whose first mathematics class is calculus I or higher had improved graduation rates compared to those whose first mathematics class was precalculus or lower. 
Second, an institution where many STEM majors started with a less rigorous first mathematics class (statistics, School B) had higher graduation rates for some STEM majors than an institution with traditional math requirements (School A). 
    These observations align with others' findings that adjustments to math requirements can improve student success.
In one example, a calculus course designed specifically for life sciences students significantly reduced the DFW rate \cite{Eaton2017-sj}. 
In another example, a contextualized two semester mathematics curriculum for chemistry, life sciences, and physics students improved both learning and interest in the subject matter \cite{OLeary2021-qu}. 
    
   As an alternative to customized calculus courses for specific majors, schools could consider adding statistics as their first mathematics requirement for many STEM majors.
   Students may be more likely to succeed in statistics as their first math class, thus generating academic momentum and giving them the opportunity to adjust to the challenges of college-level academic work in STEM before tackling more difficult classes. 

Another approach could be the implementation of new supporting classes that can be opened mid-semester for those students who are in the danger of failing in a precalculus or calculus course.
This allows students who are struggling in calculus I to withdraw from the course and still earn credits and maintain full-time status. 

Hofstra University on Long Island (NY) has improved student success in math by directing students who place into precalculus or who are doing poorly in calculus I to take specific  computer-aided modules (a program called CAMCLE) \cite{camcle2024}.  Students work semi-independently to practice both precalculus and early calculus skills before reattempting calculus the following semester. Among the 137 students who took calculus I after CAMCLE (which not all students do) 80\% passed in their first attempt and 86\% had passed by their second attempt, compared to about a 60\% pass rate in calculus I overall (David Wayne, personal communication, November 2024)
    \item[(3)] 
    Success in initial biology and chemistry courses was consistently highly ranked as relevant to graduation success of biology majors and of STEM majors in general. 
    
    This result emphasizes 
    that departments should pay attention to biology and chemistry classes, in addition to required mathematics classes, when working to improve student retention and graduation rates. 
    These classes are taken in the first two years of college, for biology and chemistry majors, and typically there is little possibility to delay them. 
    There are several strategies that could be adopted to attempt to improve success rates in introductory biology and chemistry courses.  
    One  change would be to expand a two-semester course sequence to three semesters in order to reduce the amount of course content in each course \cite{doi:10.1021/acs.jchemed.1c00387}. 
    Some schools add a pre-course requirement for those who are not college ready \cite{janice_m__bonner_2009}.
 
    However, these changes are often not adopted because they result in increasing the number of credits and/or time needed to graduate.  
    Various course redesigns have been proposed for introductory biology and chemistry courses. 
    These include instituting student-centered redesigns which was shown to decrease DFW rates \cite{Ueckert2011-pj} and focusing on core concepts (``Big Ideas") \cite{doi:10.1187/cbe.21-10-0301} in biology %\lh{please explain what this means} 
    to reduce course content and enable increased active learning \cite{Freeman2014-mo,doi:10.1187/cbe.16-09-0269,doi:10.1187/cbe.20-12-0289}.     Another proposal is to combine two introductory biology and two chemistry courses into four integrated courses \cite{https://doi.org/10.1002/bmb.21565}.
    Finally, some schools have created learning centers focused on STEM courses \cite{redd2016}.
    There are not yet published results on the efficacy of many of these approaches.

    \item[(4)] For the most part, demographic factors were not associated with graduation success or on-time graduation.  
    Identifying as Asian was strongly associated with graduation in 4 years vs.\ longer at both schools and at graduation anywhere vs.\ never graduating at School A.
    Female gender was associated with earlier graduation at both schools (Figure \ref{fig:demographics}).
    The very small numbers of records with certain attribute combinations may have made it difficult to find true associations or resulted in artificially inflated rankings for certain attributes. 

\end{itemize}

Limitations in this work include the following:
\begin{itemize}

\item Our data set is limited to two schools, which share many similarities.
Thus it is quite possible that our findings will not generalize to a broader range of different schools with different student populations.
\item We examined a limited set of student attributes that does not include many factors known to affect college outcomes, such as high school preparation and psychosocial factors. 
Many of these factors are not easily influenced by colleges and information on them was often difficult to obtain.
\item We did not examine whether intersecting identities might show different patterns of association with success outcomes. 
This would be feasible to do with the $D$-basis approach and we plan to look at this in the future.
\item The tracking data for students who left the original institution are not perfect so we do not have complete data on all students.
\item The sample size for some groups of students was relatively small.
\item Conversion of data to binary form for $D$-basis analysis led in some cases to loss of information, e.g., course grades were ``binned" so that B-, B, and B+ were all combined into a single ``B" category, and the binned grades were used in the $D$-basis rather than the actual grades. 

\end{itemize}

\section{Declarations}
\subsection{Availability of data and materials}
The datasets used and/or analysed during the current study are available from the corresponding author on reasonable request. The $D$-basis code is developed and made publicly available in github: \url{https://gitlab.com/npar/dbasis/-/tree/master}. Additional files and code for this project are located in the branch: \url{https://gitlab.com/npar/dbasis/-/tree/stem-work/analysis}.

%\subsection{Competing interests}
%The authors declare that they have no competing interests.
\subsection{Funding}
This research was funded by the National Science Foundation. 
This report is based upon work supported by the National Science Foundation under Grant Nos. 1919614 and 2121495. 
Any opinions, findings, and conclusions or recommendations expressed in this material are those of the author(s) and do not necessarily reflect the views of the National Science Foundation.
\subsection{Authors' contributions}
K.A., J.T.B., G.Z.E., L.H., and J.V.K. conceived and designed the study, acquired, analyzed, and interpreted the data, and drafted and revised the work. S.H. contributed to the data analysis and drafting of the text. M.K. and S.S. contributed to the data analysis. R.K.N. contributed to data analysis and drafting and revision of the work. O.S. created new software used in the work and contributed to data analysis.
\subsection{Acknowledgements}
We thank the staff of the Institutional Research Offices at Schools A and B for providing student data. We thank student, Wenxin Liu, for her contributions to early data analysis. We thank the (STEM)$^2$ Network (led by Jessica Santangelo, Hofstra University, and Alison Hyslop, St. John's University) for their support and for inspiring this STEM higher education community partnership. We thank Nathalia Holzmann (Queens College) for her contributions in the early stages of this project and Anne Vazquez (St. John's University) for helpful comments on the paper.

\begin{appendices}
\section{Additional files}

\subsection{Introduction to the \textit{D}-basis algorithm}
$D$-basis belongs to the family of algorithms that are based on an analysis of \emph{association rules}. The mathematical background of the algorithm is presented in in \cite{adaricheva2017discovery}. The algorithm retrieves the implications (association rules of confidence =1 ) $S \to d$ in a table with entries 0 and 1. Here $S$ is a subset of attributes/columns and $d$ is another column (say, an indicator of student graduating in 4 years since entering higher education).
The rows of the table represent students included into the study, and the presence of any attribute is indicated by entry of 1 in the column representing the attribute.

The rule $S \to d$ \emph{holds} in the table, if all entries of 1 for the set of attributes $S$ imply an
entry of 1 in column $d$, for each row of the table ( a student in our case). If all entries in $S\cup\{d\}$ are 1 in one particular row, then we say that rule $S\to d$ is \emph{validated} in that row. Set $U$ of all rows validating this rule is called a \emph{support} of the rule. 

Unlike most approaches in mining association rules, in which some rules
are selected based on techniques measuring the rules themselves, we are
looking at the measurement of \emph{relevance} of all attributes appearing in the rules with the same  \emph{target} attribute.

In particular, we measure the frequency of any
other attribute, say, $a$ as it appears in the antecedents of the rules $S \to d$, together
with other attributes. 
Our parameter of the \emph{relevance}, which appeared first in \cite{adaricheva2015measuring}, when applying the $D$-basis to the medical data, requires the computation of the rules not only on target $d$ but also on its complement $\neg d$, which may or may not appear in the original data.

Another advantage of our approach is that we do not shy away from having
a large number of retrieved rules because they provide better representation of all
attributes and allow better comparison of attributes related to a given target. 
The top attributes were identified through testing with variations of \emph{the minimal support}, which refers to the number of observations validating the rules
connecting attributes and a target. 
In our testing the number of rows (students) in the test data ran from several hundred to several thousand, and we ran the test on minimal support = 10, which would represent between 0.01-1\% of entries of the table.

\subsection{Computation of the relevance of \textit{D}-basis output}

Here we describe a new parameter useful for evaluating data from $D$-basis which we call ``relevance". 
For a fixed column $d$ and any other column $a$, one can compute the total support
of all rules $S\to d$ such that $a$ is in $S$. This parameter shows the frequency that
$a$ appears in implications targeting $d$. 
The algorithm can also compute a similar
frequency of $a$ when targeting $\neg d$, i.e., an additional column where all entries in
$d$ are switched. 
The ratio of the two frequencies gives the \emph{relevance} of attribute
$a$ to $d$ and is denoted $rel_d(a)$.

The higher $rel_d(a)$,
the more frequently attribute $a$ appears in set $S$ for rules $S\to d$ compared to
rules $S \to \neg d$. 
All attributes different from $d$, therefore, may be ranked by the
relevance with respect to $d$. Our method could investigate the attributes
with highest ranks with respect to $d$ =``graduation in 4 years" or with respect to $d$ =``student retained from year 1 to 2". 
Then all attributes of the table can be ranked in their relevance to a
fixed attribute $d$.
 
 Let us give more precise definition how the relevance of attribute $a$ with respect to target attribute $d$ is computed. 
For each attribute $a \in S\setminus{d}$, the important parameter of relevance of this attribute to $d \in X$ is the parameter of \emph{total support}, computed with respect to any set of rules/basis $\beta$ of association rules describing the table (in our case, it is the portion of the $D$-basis which only includes rules of requested minimum-support at least $\delta$):
\[
\tsup_d(a)=\Sigma \left\{\frac{|sup(X)|}{|X|}:  a\in X, (X\rightarrow d)\in \beta \text{ and } |sup(X)|\geq\delta\right\} .
\]

Thus $\tsup_d(a)$ shows the frequency of parameter $a$ appearing together with some other attributes in implications $S\rightarrow d$ of the basis $\beta$. 
The contribution of each implication $S\rightarrow d$, where $a \in S$, into the computation of total support of $a$ is higher when the support of $X$ is higher, i.e., column $a$ is marked by $1$ in more rows of the table together with other attributes from $X$, but also when $X$ has fewer other attributes besides $a$.

Note that the high frequency of some attribute in implications when targeting attribute $d$ does not mean that same attribute will appear less frequently when targeting the \emph{negation} of $d$. 
This is why the second test is done when targeting $\neg d$, which may or may not appear in the data. 
When such attribute is absent in the data, the complementary attribute can be generated from $d$. 
Thus, $\tsup_{\neg d}(a)$ could also be computed. 

Define now the parameter of relevance of attribute $a \in X\setminus d$ to attribute $d$, with respect to basis $\beta$: 
\[
\rel_d(a) = \frac{\tsup_d (a)}{\tsup_{\neg d} (a) +1}.
\]

The highest relevance of $a$ is achieved by a combination of high total support of $a$ in implications $S\rightarrow d$ and low total support in implications $U\rightarrow \neg d$.
This parameter provides the ranking of all parameters $a \in X\setminus d$.

Typically, we use a threshold $t=\frac{sup(d)}{sup(\neg d)}$ to separate the attributes that are positively relevant to $d$, i.e., $\rel_d(a) \geq t$, from the rest of attributes.

 %\ka{We may provide an example coming from the output of $D$-basis here to explain the computation of the relevance.}

\begin{example}
\end{example}
The following implication was retrieved by the algorithm from School A FT students, when we targeted attribute $d=11$, ``Student graduates (no time restriction to graduation)".  One of about 75,000 implications with minimum support $\geq 10$ that appeared in the output looked as follows:\\

(43, 60, 152, 210) $\to$ 11 ; RealSupport = 48; rows = 23, 62, 78, 80, 109, 110, 118, ...\\

Here the attributes on the left side of implication mean:

43: ``race=white"

60: ``Initial major STEM"

152: ``Gen Chemistry 1 Grade = B"

210: ``Cumulative 2d term GPA is between 3.3-3.6"\\

List of rows for the RealSupport means that for the students in the data numbered   23, 62, 78, 80, 109, 110, 118 etc (48 total), all four attributes were marked as present, and the students was also marked with attribute 11, i.e., graduated. 

The fact that this implication was retrieved also means that for any other student, besides those 48 where it was validated, at least one of attributes 43, 60, 152, 210 is not present. Thus, implication acts as it is logically defined: for each student either one of the premise attributes fails, or the conclusion (11) holds.

For each of attributes $a=$ 43, 60, 152, 210 this implication produces the following contribution into $\tsup_{11}(a)$ : $\frac{48}{4}=12$. The totals of $\tsup_{11}(a)$ are then combined across all implications, where $a$ appears. For example, $\tsup_{11}(43)=7817.83$ and $\tsup_{11}(60)=3115.71$.\\

To continue toward the computation of $rel_{11}(a)$ for one of the attributes we just considered, say, $a=43$, we also run $D$-basis with the target column $\neg d=13$, which encodes the attribute ``Student never graduated". This is the complement of $d=11$: each student who has one of these attributes does not have another, and vice versa.

For example, the same attributes  43 and 60 appeared in one of implications in the output for $\neg d=13$:\\

(1, 4, 41, 43, 60, 116) $\to$ 13 ; Real Support = 11; rows = 431, 650, 759, 860, 903, 917, 1056, 1443, 1513, 1723, 1794\\

As a result, $\tsup_{13}(43)$ and $\tsup_{13}(60)$ will have an addend for this implication: $\frac{11}{6}=1.83$. Computing the sum across all implications that have attribute 43 or 60, we will get $\tsup_{13}(43)=34.87$ and $\tsup_{13}(60)=217.35$. According to this number, attribute 60 appears more frequently in all implications of minimal support $\geq 10$ for non-graduating students.\\

Using two numbers $\tsup$ computed for attribute $a=43$, when targeting $d=11$ and $\neg d=13$, we come up with relevance of $43$ for the graduation($d=11$) :
\[
\rel_{11}(43) = \frac{\tsup_{11} (43)}{\tsup_{13} (43) +1}=\frac{7817.83}{34.87+1}\approx 217.9
\]

Similar computation for $a=60$ gives us
\[
\rel_{11}(60) = \frac{\tsup_{11} (60)}{\tsup_{13} (60) +1}=\frac{3115.71}{217.35+1}\approx 14.2
\]

In particular, attribute $a=43$ has a higher ranking than $a=60$ in the relevance to $d=11$.

The threshold in School A data: $t=\frac{sup(11)}{sup(13)}=\frac{1488}{332}=4.5$, thus, only attributes with relevance higher than this number should be considered as relevant for graduation. In particular, both attributes $a=43$ and $a=60$ are such.

%\ka{Adding a comment about unnecessarily elevated relevance values}

In some of our tests we may get $\tsup(\neg d)=0$, because the minimal support that we request rules out all found implications. We ran our tests on minimum support = 10. Therefore, if implication was validated  for 9 or fewer students, it would not appear for the calculation. This would make the denominator of relevance the smallest possible (=1).  Thus, relevance itself gets inflated compared to a similar test done on a smaller minimal support when $\tsup(\neg d)>0$.

The remedy to this situation is to produce several relevance rankings on different levels of minimal support and eliminate attributes $d$ from the top level if they only show up there due to $\tsup(\neg d)=0$.

This is equivalent to discarding statistical results obtained on small groups of population.

\subsection{Analysis of graduation in 4 years versus later graduation}

\begin{figure}[!]
    \centering
    \includegraphics[width=\textwidth]{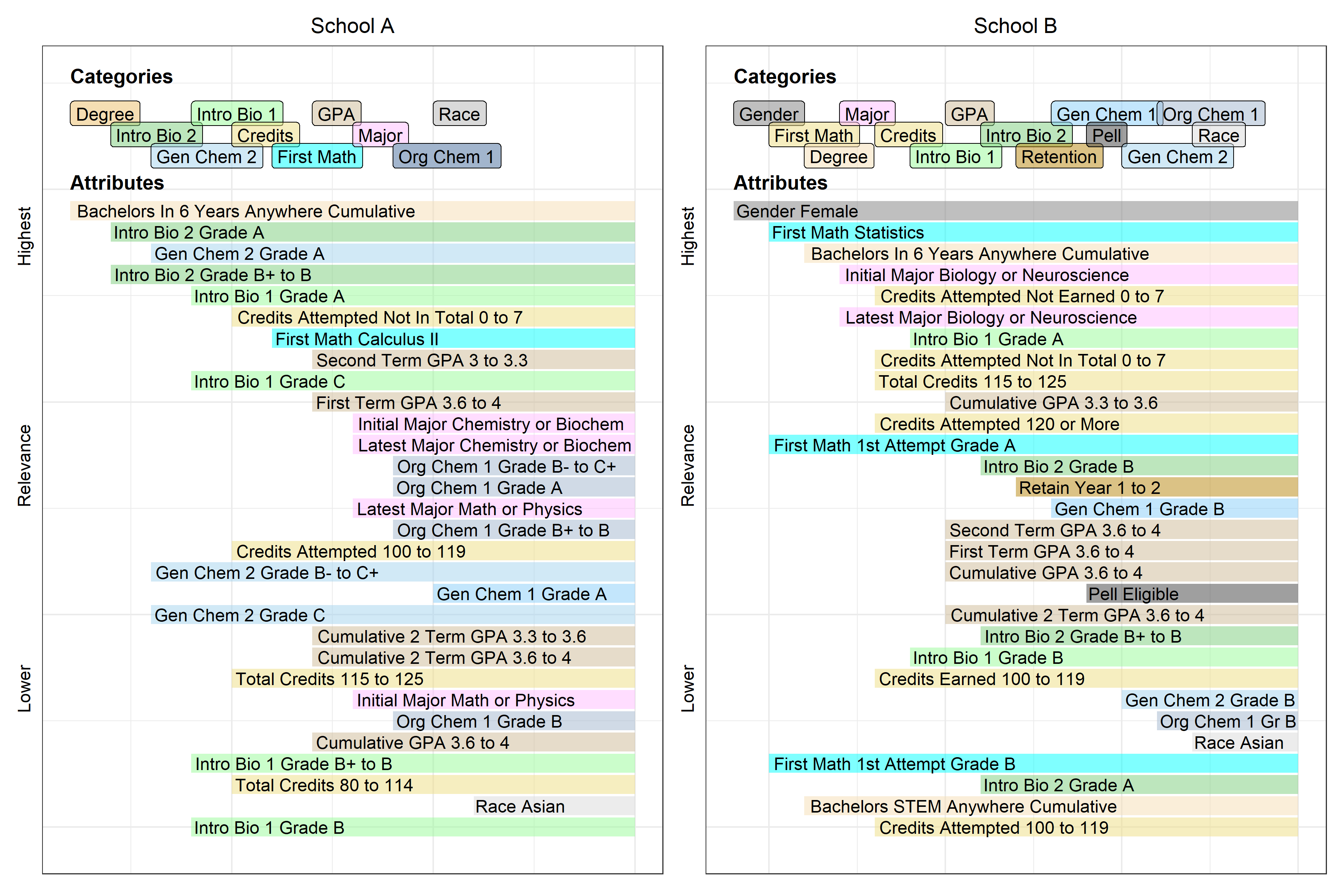}
    \caption{Top 30 attributes (out of 206) for graduation in 4 years vs.\ graduation in more than 4 years by $D$-basis analysis.  Categories of related attributes are represented in the header as rectangles. Attributes are represented as bars and are displayed in order of relevance with top-ranked at the top. Attribute bars in the same category share the same left-hand alignment and color as the corresponding category rectangle. For full list, see ( \url{https://tinyurl.com/2j3hx4yz/SchoolA_FTFT_STEM-ByRelevance_col6-col9-FullNames.csv} and \url{https://tinyurl.com/2j3hx4yz/SchoolB_FTFT_STEM-ByRelevance_col6-col9-FullNames.csv})}
%\lh{fix this when we decide. Files are %SchoolA\_FTFT\_STEM-ByRelevance\_col6-col9-%FullNames.csv and SchoolB\_FTFT\_STEM-%ByRelevance\_col6-col9-FullNames.csv })}
    \label{fig:RankedAttributesCol6vsCol9WHeaders-7-23-24}
\end{figure}

 An additional part of our analysis looked at factors correlated with shorter vs.\ longer graduation times. 
 We compared those FT STEM students who graduated in 4 years or less and those who graduated in more than 4 years, regardless of whether they graduated from their original institution or from somewhere else or whether they graduated with STEM or non-STEM degrees (although they must have had a STEM major at some point to be included in our analysis). 
The top 30 attributes by relevance for the two schools are displayed in Figure \ref{fig:RankedAttributesCol6vsCol9WHeaders-7-23-24}.

At both School A and School B, the following attributes were highly associated with $d=$ ``graduation within 4 years" and $\neg d=$ ``graduation more than 4 years": 

\begin{itemize}
    \item High overall GPA (3.3 or above).
    \item High GPA (3.6 or above) during the first two semesters.
    \item Good grades in introductory biology courses.
\end{itemize}

At the same time, the following attributes were highly associated with $\neg{d}=$ ``graduation in more than 4 years" (not shown):

\begin{itemize}
    \item Low GPA (2.0 or less), both in the first 2 semesters and overall.
    \item More than 14 credits attempted but not earned. 
    \item 39 or fewer credits attempted or 79 or fewer total credits. 
    \item Not being retained between year 1 and 2.
\end{itemize}

Besides these shared trends across both schools when targeting graduation in four years, there were also some differences between School A and School B.  
\begin{itemize} 
\item Chemistry grades: At School A high grades in organic chemistry were  associated with graduating in four years. At School B, earning a B in chemistry 1, 2, and organic chemistry were above the threshold.  
\item Major: At School A an initial or final major in chemistry, biochemistry, math or physics were above the threshold.  At School B, the majors associated with graduating in four years were biology and neuroscience.

\item Math class:  At School B, taking statistics as the first math class was ranked second out of all attributes and earning an A or B in the first math class both ranked in the top thirty attributes.  At School A, having calculus II as the first math class ranked seventh.  Math grades at School A were not associated with graduation in 4 years here.

\end{itemize}

Across both schools, when we  test alternate outcome and target graduation in {\it{more than 4 years}}, only about 30 attributes were found above the relevance threshold, indicating that students who graduate in more than 4 years are a less uniform group in their attributes than those who graduate in 4 years or less.
The high ranking of GPA factors for both schools suggests that doing poorly in first year classes is a consistent predictor of delayed graduation, if a student graduates at all.
Additionally, transferring to a different university and receiving a bachelor's degree elsewhere were highly ranked factors for delayed graduation.
This observation aligns with the factors outlined above: students who failed out of their first year but still graduated likely managed to do so because they transferred to another institution that was a better fit for them.

\begin{table}[h!]
\centering
\renewcommand{\arraystretch}{1.2}
\begin{tabular}{lccc|ccc}
   \toprule
   & \multicolumn{6}{c}{Time to Graduation} \\
   & \multicolumn{3}{c}{School A} & \multicolumn{3}{c}{School B} \\
   \cmidrule(lr){2-4} \cmidrule(lr){5-7}
   & 4 yr & 4-6 yr & $>$ 6 yr & 4 yr & 4-6 yr & $>$ 6 yr \\
   \midrule
   Total number of graduating\\ FT STEM students & 965 & 427 & 97 & 618 & 169 & 26 \\
   Percentage graduating from\\ original institution & 82\% & 53\% & 12\% & 94\% & 79\% & 23\% \\
   Percentage graduating\\ elsewhere & 18\% & 47\% & 88\% & 6\% & 21\% & 77\% \\
   \bottomrule
\end{tabular}
\caption{Graduation at the original institution or elsewhere as a percentage within the\\ groups of STEM students graduating in 4 years, between 4 and 6 years, and in more\\ than 6 years.}
\label{tab:InstGradA}
\end{table}

A more detailed analysis of FT STEM students who graduate (Table \ref{tab:InstGradA}) shows a clear trend: as the time to graduation increases, the percentage of those who graduate somewhere else also increases.  It may be surprising, therefore, that while not in the top thirty, still above the threshold, out of 218 attributes for graduation on 4 years vs.\ later the rankings of relevance for changing major at School A were STEM-to-STEM at rank 41 and STEM-to-non-STEM ranked 53, while no change of major had rank 59. 
At School B the picture was different: no change of major had rank 44 and STEM-to-STEM had rank 96.  None of the change of major attributes showed relevance to later graduation at either school. 
\begin{table}[h!]
\centering
\renewcommand{\arraystretch}{1.2}
\begin{tabular}{lccc|ccc}
   \toprule
   & \multicolumn{6}{c}{Time to Graduation} \\
   & \multicolumn{3}{c}{School A} & \multicolumn{3}{c}{School B} \\
   \cmidrule(lr){2-4} \cmidrule(lr){5-7}
   Switch of Major & 4 yr & 4-6 yr & $>$ 6 yr & 4 yr & 4-6 yr & $>$ 6 yr \\
   \midrule
   \textbf{Total number of graduating } & 965 & 427 & 97 & 618 & 169 & 26 \\
   \textbf{FT STEM students} & & & & & & \\
   \addlinespace
   No major change & 35\% & 42\% & 60\% & 56\% & 38\% & 35\% \\
   STEM-to-non-STEM & 34\% & 23\% & 16\% & 21\% & 27\% & 31\% \\
   STEM-to-STEM & 12\% & 13\% & 6\% & 6\% & 9\% & 19\% \\
   non-STEM-to-STEM & 19\% & 22\% & 18\% & 16\% & 26\% & 15\% \\
   \bottomrule
\end{tabular}
\caption{Switch of major as a percentage within the groups of STEM students\\ graduating in 4 years, between 4-6 years, and in more than 6 years.}
\label{tab:SwitchInGradA}
\end{table}

Looking at the percent of students in each path of major graduating at a particular time, we observe that at School A, changing major was correlated with \textit{faster} graduation (Table \ref{tab:SwitchInGradA}).
STEM students who did  not change their major represent a higher proportion of graduates at longer times to graduation (Table \ref{tab:SwitchInGradA}, 2nd row).
Confirming this, the highest percentage of students switching from STEM to non-STEM was among 4-year graduates (Table \ref{tab:SwitchInGradA}, 3rd row), and the percentage diminished among those who took longer to graduate.
Weaker but similar trends were found for those who switch between STEM majors and for those who change from non-STEM (usually undecided) to STEM (Table \ref{tab:SwitchInGradA}, 4th and 5th rows).
For those who graduated in 4 years, 56\% of these students did not change their major at School B but only 35\% at School A kept the same major.
For the students at School A who graduated from their \emph{original institution}, the percentage of those who did not switch majors is even smaller.
In contrast, for students who started at School A and graduated in 4 years from \emph{other} institutions, about 70\% did not change their major.
The percentage of such students is comparable at School B; however the number of such students are much smaller.
For students graduating between 4 and 6 years, at  both schools students who did not change their major now accounted for about 40\%, with the other students fairly evenly divided between switching from STEM, or into STEM. Thus it seems a larger subgroup of students in School A switch their major early and graduates in the same school, while in School B the majority of students graduating in 4 years stay with their original major.  The observation at School A does not support the interpretation that a change of majors always causes a longer path to graduation, at least for students who had a STEM major at some time in college. Instead, successful students who switched from STEM to a non-STEM major must have made this change early in their collegiate career in order to graduate within 4 years.
\end{appendices}

\bibliographystyle{alpha}
\bibliography{LIGrad}

\end{document}